\documentstyle[preprint,aps]{revtex}

 \tightenlines

\begin{document}

\title{Fourth Order Theories Without Ghosts\footnote{hep-th/0001115, 
January 19, 2000}}

\author{\normalsize{Philip D. Mannheim} \\
\normalsize{Department of Physics,
University of Connecticut, Storrs, CT 06269} \\
\normalsize{mannheim@uconnvm.uconn.edu} \\
\normalsize{and} \\
\normalsize{Aharon Davidson} \\
\normalsize{Physics Department,
Ben Gurion University of the Negev, Beer-Sheva 84105, Israel} \\
\normalsize{davidson@bgumail.bgu.ac.il}}

\maketitle

\begin{abstract}
Using the Dirac constraint method we show that the pure fourth-order
Pais-Uhlenbeck oscillator model is free of observable negative norm states. 
Even though such ghosts do appear when the fourth order theory is coupled 
to a second order one, the limit in which the second order action is 
switched off is found to be a highly singular one in which these states 
move off shell. Given this result, construction of a fully unitary, 
renormalizable, gravitational theory based on a purely fourth order action 
in 4 dimensions now appears feasible.
\end{abstract}

\bigskip

As a theory conformal gravity is somewhat enigmatic. With it being based on a 
local invariance (viz. $g_{\mu \nu}(x)\rightarrow 
e^{2\alpha(x)}g_{\mu \nu}(x)$) and with it possessing a gravitational action 
(viz. $I_W=-\alpha_g \int d^4x (-g)^{1/2} C_{\lambda\mu\nu\kappa} 
C^{\lambda\mu\nu\kappa}$ where $C^{\lambda\mu\nu\kappa}$ is the conformal Weyl 
tensor) whose coupling constant $\alpha_g$ is dimensionless, it has a structure 
which is remarkably similar to that which underlies the other fundamental 
interactions. Moreover, not only is the conformal gravitational theory power 
counting renormalizable in four spacetime dimensions, as a quantum theory it 
turns out to even be asymptotically free \cite{Tomboulis1980}, while as a 
classical theory it has \cite{Riegert1984a,Mannheim1994} the confining 
$V(r)=-\beta/r+\gamma r/2$ as its exact classical potential, to thus reproduce 
some of the most desirable properties of non-Abelian gauge theories. And 
further, through use of this linear potential the theory has been found capable 
\cite{Mannheim1997} of accounting for the perplexing galactic rotation curves 
without the need to introduce any galactic dark matter, with the cosmology 
associated with the theory being found \cite{Mannheim1998} to be free of any 
flatness, horizon, universe age or cosmological dark matter problem, with it 
being a cosmology which naturally (i.e. without fine tuning) explains the 
recently detected cosmic acceleration, and which, through its underlying scale 
invariance, naturally solves \cite{Mannheim1999} the cosmological constant 
problem. Despite all of these attractive features, standing on the debit side 
is a potential ghost problem which could render the theory non-unitary. 
Specifically, being a fourth order theory it has a $D(k^2)=1/k^4$ propagator, a 
propagator which by being rewritten as the $M^2 \rightarrow 0$ limit of 
$D(k^2,M^2)=[1/k^2-1/(k^2+M^2)]/M^2$ can immediately be seen as having 
altogether better ultraviolet behavior than the familiar $1/k^2$ propagator of 
the standard (non-renormalizable) second order gravitational theory, but which 
achieves this at the price of a possible negative norm ghost state. Prior to 
actually taking the $M^2 \rightarrow 0$ limit, the $D(k^2,M^2)$ propagator may 
be associated with the wave equation $\Box (\Box - M^2)\phi=0$ ($\Box = 
\partial^{\mu}\partial_{\mu}$), and is thus associated not with a pure fourth 
order $\Box^2\phi=0$ theory, but rather with a theory which contains both 
fourth and second order gravitational components; with it actually being this 
hybrid theory which is in fact the one which has explicitly been shown 
\cite{Stelle1977} to have a ghost. However, since the $M^2 \rightarrow 0$ limit 
of $D(k^2, M^2)$ is a singular one, one cannot immediately conclude that the 
pure fourth order theory itself necessarily also has a ghost, and in fact we 
shall show (within the simplified fourth order model first considered by Pais 
and Uhlenbeck \cite{Pais1950}) that in this limit the ghost states then 
actually decouple from the physical states and do not then lead to a 
non-unitary S-matrix. 

With a $g_{\mu\nu}=\eta_{\mu\nu} +h_{\mu\nu}$ linearization around flat 
spacetime reducing the general conformal gravity rank two gravitational tensor 
$-(-g)^{-1/2}\delta I_W / \delta g_{\mu \nu}=2\alpha_g W^{\mu \nu}=
T^{\mu\nu}/2$ to $W^{\mu \nu}=\Pi^{\mu \rho} 
\Pi ^{\nu \sigma}K_{\rho \sigma}/2- \Pi^{\mu \nu} 
\Pi ^{\rho \sigma}K_{\rho \sigma}/6$ where $K^{\mu \nu}=h^{\mu \nu}-
\eta^{\mu \nu} h^{\alpha}_{\phantom{\alpha}\alpha}/4$ and $\Pi^{\mu \nu}
=\eta^{\mu \nu} \partial^{\alpha}\partial_{\alpha}-
\partial^{\mu}\partial^{\nu}$, we find that in the conformal 
gauge\footnote{Viz. a gauge condition which is left invariant under 
$g_{\mu \nu}(x)\rightarrow e^{2\alpha(x)}g_{\mu \nu}(x)$.} 
$\partial_{\nu}g^{\mu \nu}-g^{\mu\sigma}g_{\nu\rho}\partial_{\sigma}g^{\nu 
\rho}/4=0$ the free space gravitational wave equation reduces to $\alpha_g 
\Box^2K^{\mu\nu}=0$, a pure fourth order wave equation in which the tensor 
indices nicely decouple.\footnote{With these linearized equations being found 
to be diagonal in the independent degrees of freedom, we note in passing that 
conformal gravity would thus appear to have a good initial value problem.} With 
this linearized wave equation having solutions \cite{Riegert1984b} of the form
$[a_{\mu \nu}+b_{\mu \nu}(n\cdot x)]e^{ik \cdot x}$ where $n_{\mu}=(1,0,0,0)$,
we see that the key aspect of this linearization is the presence of runaway
solutions which grow linearly in time. To explore the implications of such
temporal runaways great simplification is achieved if we ignore both the 
spatial dependence and the tensor indices and study the dynamics associated 
with the embryonic Pais and Uhlenbeck fourth order Lagrangian
\begin{equation}
L=\gamma\ddot{q}^2/2-\gamma(\omega_1^2+\omega_2^2)\dot{q}^2/2
+\gamma\omega_1^2\omega_2^2q^2/2
\label{1}
\end{equation}
instead. With this Lagrangian having equation of motion 
\begin{equation}
d^4q/dt^4+(\omega_1^2+\omega_2^2)\ddot{q}+\omega_1^2\omega_2^2 q=0,
\label{2}
\end{equation}
we see that for unequal frequencies (where the solution is
$q(t)=a_1e^{-i\omega_1t}+a_2e^{-i\omega_2t}$) the theory is analogous to
coupled fourth and second order gravitational theories, while for equal
($\omega=\omega_1=\omega_2$) frequencies (where the solution is 
$q(t)=c_1e^{-i\omega t}+c_2te^{-i\omega t}$) the theory is analogous to a 
pure fourth order theory all on its own. We shall thus quantize the theory 
associated with the unequal frequency Eq. (\ref{1}) and then study what 
happens when the equal frequency limit is taken. Since the Lagrangian 
of Eq. (\ref{1}) contains higher order time derivatives we will have to 
quantize it using the method of Dirac constraints \cite{Dirac1964}.

To bring Eq. (\ref{1}) into a form appropriate for canonical quantization we 
introduce a new variable $x(t)$ and a Lagrange multiplier $\lambda(t)$ 
according to
\begin{equation}
L=\gamma\dot{x}^2/2-\gamma(\omega_1^2+\omega_2^2)x^2/2+
\gamma\omega_1^2\omega_2^2q^2/2 +\lambda(\dot{q}-x).
\label{3}
\end{equation}
With this new Lagrangian possessing three coordinate variables, we must 
introduce three canonical momenta, which we find to be given by  
$p_x=\partial L/\partial
\dot{x} =\gamma \dot{x}$, $p_q=\partial L/\partial \dot{q}=\lambda$ and 
$p_{\lambda}=\partial L/\partial \dot{\lambda}=0$. Thus we set 
$p_q=\gamma \dot{x}$ and introduce two primary constraint functions 
$\phi_1=p_q-\lambda$, $\phi_2=p_{\lambda}$, and replace the canonical
Hamiltonian $H_c=p_x\dot{x}+p_q\dot{q}+p_{\lambda}\dot{\lambda}-L$ by
$H_1=H_c+u_1\phi_1+u_2\phi_2$ where
\begin{equation}
H_1=p_x^2/2\gamma+\gamma(\omega_1^2+\omega_2^2)x^2/2 
-\gamma\omega_1^2\omega_2^2q^2/2+\lambda x
+u_1(p_q-\lambda)+u_2p_{\lambda}.
\label{4}
\end{equation}
On introducing the canonical equal time Poisson brackets $\{x,p_x\}=\{q,p_q\}=
\{\lambda,p_{\lambda}\}=1$, we find that both of the
quantities $\{\phi_1,H_1\}=
\gamma\omega_1^2\omega_2^2 q-u_2+\phi_1\{\phi_1,u_1\}+\phi_2\{\phi_1,u_2\}$
and $\{\phi_2,H_1\}=-x+u_1+\phi_1\{\phi_2,u_1\}+\phi_2\{\phi_2,u_2\}$ will
vanish weakly (in the sense of Dirac) if we set $u_1=x$,
$u_2=\gamma\omega_1^2\omega_2^2 q$, thus enabling us to define a new 
Hamiltonian
\begin{equation}
H_2=p_x^2/2\gamma+\gamma(\omega_1^2+\omega_2^2)x^2/2 
-\gamma\omega_1^2\omega_2^2q^2/2+p_qx 
+\gamma\omega_1^2\omega_2^2qp_{\lambda}.
\label{5}
\end{equation}
For this Hamiltonian $\{\phi_2,H_2\}=\{p_{\lambda},H_2\}$ is found to be zero,
but $\{\phi_1,H_2\}$ is found to take the non-zero value
$\gamma\omega_1^2\omega_2^2p_{\lambda}$. Hence, finally, if we  now set
$p_{\lambda}=0$, the resulting algebra associated with the four-dimensional
$q,p_q,x,p_x$ sector of the theory will now (as befits a fourth
order theory) be closed under commutation, with
the requisite classical Hamiltonian then being given by
\begin{equation}
H=p_x^2/2\gamma+p_qx+\gamma(\omega_1^2+\omega_2^2)x^2/2 
-\gamma\omega_1^2\omega_2^2q^2/2,
\label{6}
\end{equation}
and with the requisite equal time Poisson bracket relations 
which define the classical theory being given by
\begin{eqnarray}
\{x,p_x\}=1,~~\{q,p_q\}=1,~~\{x,H\}=p_x/\gamma,
\nonumber \\
\{q,H\}=x,~~\{p_x,H\}=-p_q-\gamma(\omega_1^2+\omega_2^2)x,
~~\{p_q,H\}=\gamma\omega_1^2\omega_2^2q. 
\label{7}
\end{eqnarray}
The canonical equations of motion thus take the form    
\begin{equation}
\dot{x}=p_x/\gamma,~~\dot{q}=x,~~\dot{p}_x=-p_q-\gamma(\omega_1^2
+\omega_2^2)x,~~\dot{p}_q=\gamma\omega_1^2\omega_2^2q, 
\label{8}
\end{equation}
to then enable us to both recover Eq. (\ref{2}) and make the identification 
$x=\dot{q}$ in the solution, with the Legendre transform 
$L=p_x\dot{x}+p_q\dot{q}-H$ reducing to Eq. (\ref{1}) upon imposition of 
these equations of motion. The Hamiltonian $H$ of Eq. (\ref{6}) is thus the 
requisite one for the fourth order theory.

In constructing the quantum theory associated with Eq. (\ref{6}) it is 
important to distinguish between the general Hamiltonian $H$ as defined by Eq. 
(\ref{6}) and the particular value $H_{stat}$ that it takes in the stationary 
path in which the equations of motion of Eq. (\ref{8}) are imposed, with the 
great virtue of the Dirac procedure being that it allows us to define a 
Hamiltonian $H$ which takes a meaning (as the canonical generator used in Eq. 
(\ref{7})) even for non-stationary field configurations (i.e. even for 
configurations for which $p_x\dot{x}+p_q\dot{q}-H$ does not reduce to
the Lagrangian of Eq. (\ref{1})). Thus, for instance, it 
is the Hamiltonian of Eq. (\ref{6}) which defines the appropriate phase space 
for the problem, so that the path integral for the theory is then uniquely 
given by $\int [dq][dp_q][dx][dp_x] exp[i\int dt (p_x\dot{x}+p_q\dot{q}-H)]$ as 
integrated over a complete set of classical paths associated with these four 
independent coordinates and momenta. As regards the stationary value that 
$H_{stat}$ takes when the equations of motion are imposed, viz.
\begin{equation}
H_{stat}=\gamma \ddot{q}^2/2-\gamma(\omega_1^2+\omega_2^2)\dot{q}^2/2
-\gamma\omega_1^2\omega_2^2 q^2/2-\gamma \dot{q} d^3q/dt^3,
\label{9}
\end{equation}
we note that not only is $H_{stat}$ time independent (even in the runaway 
solution), it is also recognized as being the Ostrogradski 
\cite{Ostrogradski1850} generalized higher derivative Hamiltonian (viz. 
$H_{ost}=\dot{q}\partial L/\partial \dot{q}+\ddot{q}\partial L/\partial 
\ddot{q}-\dot{q}d(\partial L/\partial \ddot{q})/dt -L$) associated with the 
Lagrangian of Eq. (\ref{1}) as well as being its associated time 
translation generator \cite{Bak1994}.

With the Hamiltonian of Eq. (\ref{6}) being defined for both stationary and
non-stationary classical paths, a canonical quantization of the theory can
readily be obtained by replacing ($i$ times) the Poisson brackets of Eq. 
(\ref{7}) by canonical equal time commutators. However, without reference
to the explicit structure of the Hamiltonian itself, we note first that
the identification (as suggested but not required by the equations of 
motion\footnote{In general, equal time commutators such as $[q(t),p_q(t)]=i$
can be satisfied at all times by an $[a,a^{\dagger}]=1$ commutator defined
via $q(t)=af(t)/2^{1/2}+H.c.$, $p_q(t)=ia^{\dagger}/2^{1/2}f(t)+H.c.$ with the
function $f(t)$ being arbitrary. The occupation number Fock space can thus
be defined independent of the structure of $H$, and even independent of any
interaction terms that might also be added on to the free particle
$H$. Thus in general, and as we shall explicitly see below in
particular, the dimensionality of the Fock space basis need not be the same
as that of the eigenspectrum of $H$.})
\begin{eqnarray}
q(t)=a_1e^{-i\omega_1t}+a_2e^{-i\omega_2t}+H.c.,~~
p_q(t)=i\gamma \omega_1\omega_2^2a_1e^{-i\omega_1t}+
i\gamma \omega_1^2\omega_2 a_2e^{-i\omega_2t}+H.c.,
\nonumber \\
x(t)=-i\omega_1a_1e^{-i\omega_1t}-i\omega_2 a_2e^{-i\omega_2t}+H.c.,~~
p_x(t)=-\gamma\omega_1^2a_1e^{-i\omega_1t}-
\gamma \omega_2^2a_2e^{-i\omega_2t}+H.c.
\label{10}
\end{eqnarray}
then furnishes us with a Fock space representation of the quantum-mechanical 
commutation relations $[x,p_x]=[q,p_q]=i$, $[x,q]=[x,p_q]=[q,p_x]=[p_x,p_q]=0$ 
at all times provided that
\begin{equation}
[a_1,a_1^{\dagger}]=[2\gamma\omega_1(\omega_1^2-\omega_2^2)]^{-1},~~
[a_2,a_2^{\dagger}]=[2\gamma\omega_2(\omega_2^2-\omega_1^2)]^{-1},~~
[a_1,a_2^{\dagger}]=0,~~[a_1,a_2]=0.
\label{11}
\end{equation}
Then in this convenient Fock representation  the quantum-mechanical 
Hamiltonian is found to take the form
\begin{equation}
H=2\gamma(\omega_1^2-\omega_2^2)(\omega_1^2a_1^{\dagger}a_1-
\omega_2^2a_2^{\dagger}a_2)
+(\omega_1+\omega_2)/2
\label{12}
\end{equation}
with its associated commutators as inferred from Eq. (\ref{7}) then 
automatically being satisfied. With the quantity
$\gamma(\omega_1^2-\omega_2^2)$  being taken to be positive for
definitiveness, we see that the $[a_2,a_2^{\dagger}]$ commutator is negative
definite, and with $H$ being diagonal in the $a_1,a_2$ occupation number
basis, we see that the state defined  by $a_1 |\Omega\rangle=0$, $a_2
|\Omega\rangle =0$ is the ground state of 
$H$,\footnote{Identifying $a^{\dagger}$ as the annihilator of the Fock vacuum
would eliminate negative norm states but would leave $H$ without any lower 
bound on the ground state energy.} that the states 
$|+1\rangle =[2\gamma\omega_1(\omega_1^2-\omega_2^2)]^{1/2}
a_1^{\dagger}|\Omega\rangle$, 
$|-1\rangle =[2\gamma\omega_2(\omega_1^2-\omega_2^2)]^{1/2}
a_2^{\dagger}|\Omega\rangle$ are both positive energy eigenstates 
with respective energies $\omega_1$ and $\omega_2$ above the ground state, 
that the state $|+1\rangle$ has a norm equal to plus one, but that the state 
$|-1\rangle$ has norm minus one, a ghost state. With the eigenstates of $H$
labeling the asymptotic states associated with scattering in the presence of
an interaction Lagrangian $L_I$ which might be added on to the original
Lagrangian $L$ of the theory, the action of $L_I$ would induce transitions
between asymptotic in and out states of opposite norm, with the unequal
frequency theory thus explicitly being seen to be 
non-unitary.\footnote{While it is widely believed that the quantum theory 
associated with Eq. (\ref{1}) is not unitary, it is possible that our 
demonstration here might be the first explicit constructive proof.} However, 
even though the Hamiltonian of the unequal frequency theory does have ghost 
eigenstates, since the commutation relations given in Eq. (\ref{11}) become 
singular in the equal frequency limit while both the Hamiltonian of Eq. 
(\ref{12}) and the normalized $|+1\rangle$ and $|-1\rangle$ states develop
zeroes, we will have to exercise some caution in trying to discover exactly
what happens to the ghost states when this limit is taken. And in fact,
we will find below that there are actually two possible ways to take this
limit, ways which will lead to differing energy eigenspectra even as they lead
to the same commutator algebra, with both ways being found to lead to a
unitary theory.

To explore the limit it is convenient to
introduce new Fock space variables according to $a_1+a_2=a-b$,
$a_1-a_2=2b\omega/\epsilon$ where 
$\omega=(\omega_1+\omega_2)/2$ and $\epsilon=(\omega_1-\omega_2)/2$. These 
variables are found to obey commutation relations $[a,a^{\dagger}]=\lambda$, 
$[a,b^{\dagger}]=\mu$, $[b,b^{\dagger}]=\nu$, where
\begin{equation}
\lambda=-\epsilon^2/16\gamma\omega_1\omega_2\omega^3,~~
\mu=(2\omega^2-\epsilon^2)/
16\gamma\omega_1\omega_2\omega^3,~~
\nu=-\epsilon^2/
16\gamma\omega_1\omega_2\omega^3.
\label{13}
\end{equation}
In terms of these variables $q(t)$ of Eq. (\ref{10}) can be reexpressed as
\begin{equation}
q(t)=e^{-i\omega t}[(a-b)cos~\epsilon t-
2ib\omega\epsilon^{-1} sin~\epsilon t] +H.c.,
\label{14}
\end{equation}
and thus has a well defined $\epsilon \rightarrow 0$ limit, viz. 
\begin{equation}
q(t)=e^{-i\omega t}(a-b-2ib \omega t) +H.c.,
\label{15}
\end{equation}
a limit in which the Hamiltonian of Eq. (\ref{12}) takes the form 
\begin{equation}
H(\epsilon=0)=8\gamma \omega^4 (2b^{\dagger}b+a^{\dagger}b 
+ b^{\dagger}a)+\omega,
\label{16}
\end{equation}
and in which the following commutators of interest take the form  
\begin{eqnarray}
~~[H(\epsilon=0),a^{\dagger}] 
=\omega (a^{\dagger}+2b^{\dagger}),~~[H(\epsilon=0),b^{\dagger}]=
\omega b^{\dagger},
\nonumber \\
~~[a+b,a^{\dagger}+b^{\dagger}]=2\mu,~~[a-b,a^{\dagger}-b^{\dagger}]=-2\mu,
~~[a+b,a^{\dagger}-b^{\dagger}]=0.
\label{17}
\end{eqnarray}
Thus we see that use of the $a^{\dagger},~b^{\dagger}$ operators enables us
to construct an $\epsilon=0$ theory which is well defined, with the 
$\epsilon=0$ Fock space then being built on the Fock vacuum $|\Omega\rangle$ 
defined by $a|\Omega\rangle=b|\Omega\rangle=0$ (we need to use this basis
since states built on the unequal frequency Fock vacuum  become undefined in
the $\epsilon=0$ limit). Moreover, in the
$\epsilon=0$ limit we see that the $[a,a^{\dagger}]$ and $[b,b^{\dagger}]$
commutators both  vanish, with, as we shall see below, there actually being
two ways rather than one in which the Fock space states can implement this
vanishing. However, before constructing these two ways, we note that because
the action of the Hamiltonian is to shift $a^{\dagger}$ by $2b^{\dagger}$,
and because (unlike the unequal frequency case) neither of the operators 
($a^{\dagger} \pm b^{\dagger}$) which diagonalize the Fock space basis acts as
a ladder operator for the Hamiltonian, we can anticipate that the
dimensionality of the eigenspectrum of $H(\epsilon=0)$ will  be lower than
that of $[x,p_x]=[q,p_q]=i$, $[x,q]=[x,p_q]=[q,p_x]=[p_x,p_q]=0$ commutator
algebra, i.e. lower than the dimensionality of the two-dimensional harmonic
oscillator, even though that precisely was the dimensionality of the unequal
frequency eigenspectrum.   

Because $b^{\dagger}$ does act as a ladder operator for $H(\epsilon=0)$,
it is possible to construct a one particle energy eigenstate, viz. the state
$b^{\dagger}|\Omega\rangle$, but if we try to normalize it
(our first way to realize the $\epsilon=0$ Fock space) the vanishing of the
$[b,b^{\dagger}]$ commutator would entail that this state would have to have
zero norm. On its own a zero norm state (unlike a negative norm state) does
not lead to loss of probability, but its very existence entails the existence
of negative norm states elsewhere in the theory.\footnote{The unequal 
frequency theory states $|+1\rangle \pm |-1 \rangle$ both have zero norm.}
Thus, with the state $a^{\dagger}|\Omega\rangle$ also having zero norm, we
immediately construct the states $|\pm\rangle = (a^{\dagger}\pm
b^{\dagger})|\Omega\rangle/(2\mu)^{1/2}$, states which obey $\langle
+|+\rangle =1$, $\langle -|-\rangle =-1$, $\langle +|-\rangle =0$. 
However, unlike the orthogonal positive and negative norm states $|\pm1\rangle$
associated with the unequal frequency case, this time we find that neither of
the $|\pm\rangle$ states is an eigenstate of the Hamiltonian
($H(\epsilon=0)|\pm\rangle=2\omega(|\pm\rangle +2b^{\dagger}|\Omega\rangle$)).
Thus even while the $\epsilon=0$ Fock space possesses negative norm ghost
states, this time they can only exist off shell (where they can still regulate
Feynman diagrams) but cannot materialize as on shell asymptotic in and out
states. With a similar situation being found in the two particle 
sector\footnote{The zero norm states $(b^{\dagger})^2|\Omega\rangle$ and 
$(a^{\dagger})^2|\Omega\rangle$ can be combined into positive and negative 
norm states while the state $a^{\dagger} b^{\dagger}|\Omega\rangle/\mu$ has
norm plus one.} where only $(b^{\dagger})^2|\Omega\rangle$ is an energy
eigenstate, we see that the complete spectrum of eigenstates of
$H(\epsilon=0)$ is the set of all states $(b^{\dagger})^n|\Omega\rangle$,
a spectrum whose dimensionality is that of a one rather than a
two-dimensional harmonic oscillator, even while the complete Fock space
has the dimensionality of the two-dimensional oscillator. The equal
frequency fourth order theory thus has a propagator which acts not like        
a $D(k^2,M^2)=[1/k^2-1/(k^2+M^2)]/M^2$ propagator with poles at
$k_0=\pm k,~k_0=\pm (k^2+M^2)^{1/2}$, but rather as the derivative
operator $D(k^2)=-d(k^{-2})/dk^2)$, i.e. as the derivative of a propagator 
which only has poles at $k_0=\pm k$.\footnote{A further unusual property of
the equal frequency theory is that the one particle matrix element 
$\langle \Omega| q(t) b^{\dagger} | \Omega \rangle$ is given by 
$ \mu e^{-i\omega t}$ rather than by the coefficient of the $b$ field operator
in Eq. (\ref{15}), with there thus being a mismatch between the second
quantized states and the first quantized wave functions.} Then finally, with
the free and interacting Fock spaces being the same (the Fock space can be
defined once and for all by the
$[x,p_x]=[q,p_q]=i$, $[x,q]=[x,p_q]=[q,p_x]=[p_x,p_q]=0$ commutator algebra)
and with $(b^{\dagger})^n|\Omega\rangle$ subset of Fock states serving to 
label the asymptotic in and out energy eigenstates, we see that the equal
frequency theory is unitary.   

While the $\epsilon=0$ theory is thus seen to be unitary, it is nonetheless
a somewhat unusual theory in that all of its energy eigenstates have zero
norm. Thus it would be of interest to see if we could construct an alternate
limiting theory in which the energy eigenstates have positive norm instead.
To this end we note that since we obtained the zero norm states by first
going to the $\epsilon=0$ limit and then constructing the states, an
alternate procedure would be to first construct normalized states and then
take the limit. With the algebra of the $a$ and $b$ operators already being
defined prior to taking the limit, we can thus construct a basis for the 
$\epsilon \neq 0$ Fock space via states such as (the choice $\gamma<0$ assures 
the positivity of $\lambda$ and $\nu$) the one particle 
$|1,0\rangle=a^{\dagger}|\Omega\rangle/\lambda^{1/2}$ and 
$|0,1\rangle=b^{\dagger}|\Omega\rangle/\nu^{1/2}$, the two particle 
$|2,0\rangle=(a^{\dagger})^2|\Omega\rangle/2^{1/2}\lambda$,
$|1,1\rangle=a^{\dagger}b^{\dagger}|\Omega\rangle/(\mu^2+\lambda\nu)^{1/2}$ 
and $|0,2\rangle=(b^{\dagger})^2|\Omega\rangle/2^{1/2}\nu$, and so on, 
a complete basis in which, remarkably, every single state is found to have 
positive norm, but not an orthogonal one since overlaps such as 
$\langle 1,0|0,1\rangle=\mu/(\lambda \nu)^{1/2}$ are non-zero, overlaps which 
actually become singular in the $\epsilon \rightarrow 0$ limit. Since 
this basis is complete we can use it to define a limiting procedure in which 
each of the basis states is held fixed while $\epsilon$ is allowed to go to 
zero. Then in such a limit we find that $a^{\dagger}|\Omega\rangle$ and 
$b^{\dagger}|\Omega\rangle$ both become null vectors. However, despite this, 
we cannot conclude that the creation operators 
annihilate the vacuum identically in this limit since matrix elements such as
$\langle \Omega|[a,b^{\dagger}]|\Omega\rangle = \mu $ do not vanish 
in the limit, with  product operator actions such as $a$ acting
on $b^{\dagger}|\Omega\rangle$ being singular. Instead, the creation operators 
must be thought of as annihilating the vacuum weakly (i.e. in some but not all 
matrix elements), but not strongly as an operator identity. Now since 
neither of the states $a^{\dagger}|\Omega\rangle$ or 
$b^{\dagger}|\Omega\rangle$ survives in the limit, $H(\epsilon=0)$ now has no 
one particle eigenstates at all. With the two particle states 
$(a^{\dagger})^2|\Omega\rangle$ and $(b^{\dagger})^2|\Omega\rangle$ also
becoming null in the limit, the only two particle state which is found to 
survive in this limit is the positive norm state 
$a^{\dagger}b^{\dagger}|\Omega\rangle/\mu$ (since the $[a,b^{\dagger}]$ 
commutator does not vanish), and, quite remarkably, in this limit the 
state is also found to actually become an energy eigenstate
($H(\epsilon=0)a^{\dagger}b^{\dagger}|\Omega\rangle/\mu
=3\omega a^{\dagger}b^{\dagger}|\Omega\rangle/\mu+
2\omega (b^{\dagger})^2|\Omega\rangle/\mu \rightarrow
3\omega a^{\dagger}b^{\dagger}|\Omega\rangle/\mu$). With this analysis
immediately generalizing to the higher multiparticle states as well, we see
that the only states which then survive in the limit are states of the form
$(a^{\dagger}b^{\dagger})^n|\Omega\rangle$, positive norm states which also
become energy eigenstates in the limit,
with the observable sector of the theory again being unitary, and with it 
again having the dimensionality of a one-dimensional harmonic oscillator and
not that of a two-dimensional one.\footnote{In the conformal gravity case
such a situation would correspond to there only being one observable graviton
rather than two, with the observable graviton being composite rather
than fundamental.}  We thus recognize two limiting procedures, first taking
the limit of the algebra and then constructing the Hilbert space, or first
constructing the Hilbert space and then taking the limit of the algebra, with
this latter limiting procedure being an extremely delicate one in which the
only states which survive as observable ones are composite.\footnote{A model
in which a field has a positive frequency part which annihilates the vacuum
strongly and a negative frequency part which annihilates the vacuum weakly in
a way such that only certain multiparticle states survive would appear to be a 
possible candidate mechanism for quark confinement, with quarks themselves 
then only existing off shell.} Thus, to conclude, we
see that in the equal frequency limit the quantum theory based on the
fourth order Lagrangian of Eq. (\ref{1}) is in fact unitary (in fact, 
technically, what we have shown is that the lack of unitarity of
the unequal frequency theory is simply not a reliable indicator as to the 
unitarity status of the equal frequency one), and that construction
of a fully unitary and renormalizable quantum gravitational theory in four
spacetime dimensions would now appear feasible. One of the authors (P.D.M)
wishes to thank Dr. E. E. Flanagan for useful discussions. The work of P.D.M.
has been supported in part by the Department of Energy under grant No.
DE-FG02-92ER4071400.

\end{document}